\documentclass[12pt,leqno]{article}
 \usepackage{amsmath}
 \usepackage{amsfonts,latexsym,amssymb}
\usepackage{amsthm}
 \usepackage{newtxtext}          
 \usepackage{newtxmath}
 \usepackage[T1]{fontenc}       
\usepackage[utf8]{inputenc}     
 \usepackage{varioref}
 \usepackage{enumitem}

\usepackage{graphicx}
\usepackage[all]{xy}
\usepackage[pdftex,bookmarks]{hyperref}

\hypersetup{
    pdfnewwindow=true,          
    colorlinks=true,           
    linkcolor=blue,            
    citecolor=blue,           
    filecolor=magenta,          
    urlcolor=cyan,              
    pdfauthor=D\"untsch \& Or{\l}owska,   
}
\usepackage[ocgcolorlinks]{ocgx2}

\usepackage{geometry}
\usepackage{xspace}
\usepackage[numbers,sort&compress]{natbib}

\renewcommand{\newblock}{} 
\newcommand{\Ult}{\operatorname{Ult}}
\newcommand{\Prim}{\operatorname{Prim}}
\newcommand{\klam}[1]{\ensuremath{\langle #1 \rangle}}
\newcommand{\nec}[1]{\ensuremath{[ #1 ]}}
\newcommand{\poss}[1]{\ensuremath{\klam{ #1 }}}
\newcommand{\suff}[1]{\ensuremath{[[ #1 ]]}}
\renewcommand{\frm}{\ensuremath{\mathsf{Frm}}\xspace}
\newcommand{\alg}{\ensuremath{\mathsf{Alg}}\xspace}
\newcommand{\set}[1]{\ensuremath{\{#1\}}}
\newcommand{\z}{\emptyset}
\newcommand{\df}{\mathrel{\mathop:}=}
\newcommand{\tiff}{if and only if\xspace}
\newcommand{\tand}{\text{ and }}

\newcommand{\into}{\hookrightarrow}
\newcommand{\Implies}{\ensuremath{\mathrel{\Rightarrow}}}
\newcommand{\Iff}{\ensuremath{\mathrel{\Longleftrightarrow}}}
\newcommand{\Iffdf}{\overset{\mathrm{df}}{\Longleftrightarrow}}

\newcommand{\Cf}{\ensuremath{\operatorname{{\mathsf{Cs}}}}}
\newcommand{\Cs}{\ensuremath{\operatorname{{\mathsf{Cs}}}}}
\newcommand{\Cm}{\ensuremath{\operatorname{{\mathsf{Cm}}}}}

\newcommand{\B}{\ensuremath{\mathcal{B}}\xspace}

\newcommand{\X}{\ensuremath{\mathcal{X}}}

\newcommand{\zero}{\mathbf{0}} 
\newcommand{\one}{\mathbf{1}} 

\newcommand{\tor}{\text{ or }}
\newcommand{\timplies}{\text{ implies }}

\newcommand{\comp}{\mathrel{;}}
\newcommand{\conv}[1]{#1\ensuremath{~\breve{}~}}

\newcommand{\id}{\one'}
\newcommand{\rel}{\operatorname{Rel}}

\newcommand{\da}[1]{\ensuremath{\mathop{\downarrow}#1}}
\newcommand{\ua}[1]{\ensuremath{\mathop{\uparrow}#1}}

\numberwithin{equation}{section}

\parskip=\medskipamount
\numberwithin{equation}{section}

\theoremstyle{plain}
\newtheorem{theorem}{Theorem}[section]
\newtheorem{lemma}[theorem]{Lemma}

\date{}

\begin{document}

\title{Discrete dualities for some algebras from rough sets}

\author{
Ivo D\"untsch%
 \\
{Dept of Computer Science} \\
{Brock University} \\
{St Catharines, ON, L2S 3A1, Canada} \\
{\url{duentsch@brocku.ca}}
\and
Ewa Or{\l}owska \\
{Institute of Telecommunications} \\
{Szachowa 1} \\
{04--894, Warszawa, Poland} \\
{\url{stella.ewa.orlowska@gmail.com}}
}

\maketitle

\abstract{
\noindent A discrete duality is a relationship between classes of algebras and classes of relational systems (frames) resulting in two representation theorems building on the early work of \citet{jt51}, \citet{kripke1963}, and \citet{Benthem84}. In this section we recall discrete dualities for various types of algebras arising from rough sets.
}

\section{Introduction}

In the present article we present the place of discrete duality in the landscape of knowledge and some of its examples arising in the field of rough sets. Areas of contemporary knowledge can be divided into two major groups, namely, science and humanities. Science includes areas such as mathematics and physics, humanities include literature and psychology, among others. Each of the science fields has its own methodology. Methodology of the fields in the science group can be traced back to the beginning of the age of enlightenment. In his treatise \emph{Discourse de la Methode} (Discourse on Method \cite{des1637}, translation from the French by John Veitch \cite{des1912}, the mathematician  and physicist Ren{\'e} Descartes (March 31, 1596, La Haye -- February 11, 1650, Stockholm)  presented four principles that must be followed by researchers

\begin{quote}
''The first was never to accept anything for true which I did not clearly know to be such; that is to say, carefully to avoid precipitancy and prejudice, and to comprise nothing more in my judgement than what was presented to my mind so clearly and distinctly as to exclude all ground of doubt.

The second, to divide each of the difficulties under examination into as many parts as possible, and as might be necessary for its adequate solution.

The third, to conduct my thoughts in such order that, by commencing with objects the simplest and easiest to know, I might ascend by little and little, and, as it were, step by step, to the knowledge of the more complex; assigning in thought a certain order even to those objects which in their own nature do not stand in a relation of antecedence and sequence.

And the last, in every case to make enumerations so complete, and reviews so general, that I might be assured that nothing was omitted.'' \cite[p. 15f]{des1912}
\end{quote}
The attention to rigour and strictness postulated by Descartes was an indication of a trend  leading to mathematical logic becoming  \emph{metamathematics}. Indeed, since the turn of the 17$^{\mathrm{th}}$ century many abstract concepts were defined such as that of a function. The 19$^{\mathrm{th}}$ century is the period when creation of concepts and investigation of their properties are the essence of the activities of mathematicians and logicians The seminal work by Gottlob Frege, Alfred Whitehead, Bertrand Russell, Ludwig Wittgenstein, followed in the early 20$^{\mathrm{th}}$ century by David Hilbert, Kurt G{\"o}del and Gerhard Gentzen, is the foundation of modern mathematics and logic.

Nowadays, a commonly accepted and practiced approach to formalization of mathematical theories is to define a class \alg of algebras and a class \frm of frames (relational structures) such that \alg represents the deduction tools for the given data, and \frm represents the semantics for the data. A discrete duality of \alg and \frm is a tool for proving that those two classes correctly reflect the theories under consideration. In detail, a \emph{discrete duality} between \frm and \alg holds whenever there are maps $\Cm\colon \frm \to \alg$  and $\Cf\colon \alg \to \frm$ such that for each frame $X$ from \frm the algebra $\Cm(X))$, referred to as the \emph{complex algebra} of  $X$, belongs to \alg,  and for each algebra $L$ from \alg the frame $\Cf(L)$, referred to as the \emph{canonical structure} of $L$, belongs to \frm. Furthermore, the following two representation theorems hold which show how the classes \frm and \alg are related \cite{orr15}:

\begin{itemize}
\item \textbf{Representation theorem for algebras}

Every $L \in \alg$ is embeddable into the complex algebra of its canonical structure $\Cm(\Cf(L)$.

\item \textbf{Representation theorem for frames}

Every $X \in \frm$ is embeddable into the canonical structure of its complex algebra $\Cf(\Cm(X))$.

\end{itemize}

We consider two approaches to define algebras from approximation spaces and rough sets following Pawlak's original ideas on granularity and approximation of information \cite{mod25}:
\begin{enumerate}
\item As Boolean algebras with operators induced by the equivalence relation of an approximation space. This approach leads to the variety of monadic algebras, which are the complex algebras of Kripke models for the logic \textbf{S5}. 
\item As algebras obtained by defining lattice operations on the collection of rough sets of an approximation space. This leads to, among others,  regular double Stone algebras and De Morgan algebras. 
\end{enumerate}

Below we list the definitions and conventions we follow in the rest of the paper. 

For $x \in X$, we let $R(x) \df \set{y \in X: x \mathrel{R} y}$. If $R$ is a partial order, say $\leq$, then, $R(x)$ is the principal filter generated by $x$, denoted by $\ua_\leq{x}$. More generally, if $A \subseteq X$, we let $\ua_\leq{A} \df \set{y \in X: (\exists x \in X)[x \in A \tand x \leq y]}$ be the order filter generated by $A$. Similarly we define $\da{A}$ as the order ideal generated by $A$.  We omit the subscript, if $\leq$ is understood. If $\klam{X,R}$ and $\klam{Y,S}$ are frames, a mapping $k\colon X \to Y$ is a \emph{frame embedding}, if $k$ is injective and for all $x,y \in X$,
\begin{gather}
x \mathrel{R} y \Iff k(x) \mathrel{S} k(y),
\end{gather}
that is, if $k$ preserves and reflects $R$. 

With some abuse of notation we shall usually identify algebras and frames with their universe, if no confusion can arise. Equations are assumed to be universally quantified. For unexplained concepts and notation for rough sets we refer the reader to the compendium \cite{orr15}.

\section{Monadic algebras}\label{sec:modal} 

Our first class of algebras obtained from rough sets arises directly from the Pawlak's concepts of approximation of information. Let us recall the definitions of the approximation operators of an approximation space $\klam{X,\theta}$, where $\theta$ is an equivalence relation on $X$. If $Y \subseteq X$, then
\begin{xalignat}{2}
\tag{Up}\label{Yp} \overline{Y}_\theta &\df \set{x \in Y: (\exists y \in Y)x \mathrel{\theta} y},  &&\text{Upper approximation of $Y$}, \\
\tag{Low}\label{Low} \underline{Y}_\theta &\df\set{x \in Y: (\forall y \in Y)[x \mathrel{\theta} y \Implies y \in Y]}, &&\text{Lower approximation of $Y$}.
\end{xalignat}
If $\theta$ is understood, we will omit the subscripts. Since $\theta$ is an equivalence relation, we may consider $\klam{X,\theta}$ as a Kripke frame for the modal logic \textbf{S5} with accessibility relation $\theta$. The \emph{complex algebra} $\Cm_\theta(X)$ of $\klam{X,\theta}$ is the structure $\klam{2^X,\poss{\theta}}$ , where $2^X$ is the Boolean powerset algebra of $X$ augmented by the \emph{possibility operator} $\poss{\theta}\colon 2^X \to 2^X$ defined by 
\begin{xalignat}{2}
\label{possR} \poss{\theta}(Y) &\df \set{x \in X: \theta(x) \cap Y \neq \z}.
\end{xalignat}
Its dual $\nec{\theta}$ is defined by $\nec{\theta}(Y) = -\poss{\theta}(-Y)$. It is called the \emph{necessity operator} since $\nec{\theta}(Y) = \set{x: \theta(y) \subseteq Y}$. i.e. for $x \mathrel{\theta} y$ it is necessary that $y \in Y$. These operators first appeared in the seminal work of \citet{jt51} and were later used in the context of Kripke frames of propositional logics. A collection of properties of these operators can be found in \cite[Section 1.8]{orr15}. They indeed capture the intention of the rough set approximation operators, since
\begin{align}
x \text{ is possibly in } Y &\Iff x \in \poss{Y}, \\
x \text{ is certainly in } Y &\Iff x \in \nec{Y}.
\end{align}
The general situation is described by a class of Boolean algebras with an additional operator. A \emph{possibility algebra} is a Boolean algebra $\klam{B,+,\cdot,-,\zero.\one}$ augmented by a unary operator $f$ which satisfies
\begin{xalignat}{2}
\tag{K1}\label{K1} f(\zero) & = \zero, && \text{Normality,} \\
\tag{K2}\label{K2} f(a+ b) &= f(a) + f(b), &&\text{Additivity.}
\end{xalignat}
The dual $f^\partial$ of $f$ is the operator on $B$ defined by $f^\partial(a) \df -f(-a)$. 

A possibility algebra $\klam{B,f}$ is called a \emph{monadic algebra} if it satisfies
\begin{align}
\tag{T}\label{T} a &\leq f(a), \\
\tag{4}\label{4} f(f(a)) &\leq f(a), \\
\tag{B}\label{B} f(a) &=  f^\partial(f(a)).
\end{align}
There are various equivalent axiomatizations of monadic algebras. We have chosen the axiomatization above and their names to stay close to the logical nature of the axioms. The following properties of a monadic algebra are well known, see e.g. \cite{jt51,dav54,hal56}:
\begin{lemma}
Let $\klam{B,f}$ be a monadic algebra. Then,
\begin{xalignat}{2}
a &\leq f^\partial(f(a)), && \text{The Euclidean axiom \textbf{5},} \label{A5}\\
f(a) \leq b &\Iff a \leq f^\partial(b), &&\text{$f$ and $f^\partial$ are adjoint.} \label{Adj} \\
a \cdot f(b) = \zero &\Iff f(a) \cdot b = \zero, &&\text{$f$ is self conjugate,} \label{conj}\\
f(a \cdot f(b)) &= f(a) \cdot f(b), &&\text{The cylindric axiom.}\label{df1}
\end{xalignat}
\end{lemma}
The \emph{canonical structure} $\Cf(B)$ of a monadic algebra $\klam{B,f}$ is the structure $\klam{\Ult(B), \theta_f}$ where $\Ult(B)$ is the set of ultrafilters of $B$, and $\theta_f$ is the binary relation on $\Ult(B)$ defined by
\begin{gather}\label{def:Rf}
F \mathrel{\theta_f} G \Iffdf f[G] \subseteq F.
\end{gather}
The next decisive result is well known:
\begin{lemma}\label{lem:equiv} \cite{dav54} 
If $\klam{B,f}$ is a monadic algebra, then $\theta_f$ is an equivalence relation of $\Ult(B)$. \qed
\end{lemma}
For the other direction recall the definition of the complex algebra $\Cm(X)$ of $\klam{X,\theta}$  as $\klam{2^X, \poss{\theta}}$, where $\poss{\theta}$ is defined in \eqref{possR}.
\begin{lemma}\label{lem:monadic}\cite[Theorem 3.5]{jt51} 
If $\klam{X,\theta}$ is an approximation space, then $\Cm(X)$ is a monadic algebra. \qed
\end{lemma}

We are now ready to exhibit the discrete duality between the class of monadic algebras and the class of approximation spaces. Since monadic algebras are Boolean algebras with operators in the sense of \citet{jt51}, our procedure is based on the discrete duality for such algebras as shown in \cite[Theorem 3.2.7]{orr15} which, in turn, relies on Stone's representation theorem for Boolean algebras \cite{stone_BA}. 
 
\begin{theorem}\label{thm:monad}
 \begin{enumerate}
 \item $\Cm(\Cf(B))$ is a monadic algebra, and $h\colon B \into \Cm(\Cf(B))$ defined by $h(a) \df \set{F \in \Ult(B): a \in F}$ is an embedding of Boolean algebras.
 \item $\Cf(\Cm(X))$ is an approximation space and the mapping $k\colon \Cf(\Cm(X))$ defined by $k(x) \df \ua_{\subseteq}{\set{x}}$ is a frame embedding. \qed
 \end{enumerate}
 \end{theorem}

One can relax the condition that $\theta$ is an equivalence relation and obtain corresponding discrete dualities by translating the relational condition into algebraic equations such as in \cite[Theorem 3.5]{jt51}. One can also consider more than one information relation and their interaction. An overview of these can be found in \cite{orl98a} and an algebraic investigation of the associated algebras was conducted in \cite{do_baro}.

\section{Sufficiency algebras}

While upper and lower approximation of sets given an approximation space $\klam{X,\theta}$ are natural semantic interpretations of monadic algebras, we will introduce another operator which can be naturally obtained from an approximation space. Let $R$ be a binary relation on $X$, and define the mapping $\suff{R}\colon 2^X \to 2^X$ by
\begin{gather}
\suff{R}(Y) \df \set{x: (\forall y)[y \in Y \Implies x\mathrel{R}y]} = \set{x: Y \subseteq R(x)}. \label{def:suff}
\end{gather}
for $Y \subseteq X$. 
The set $\suff{R}(Y)$ collects all those elements of $X$ whose $R$-class contains $Y$, in other words, for $x \mathrel{R} y$ to hold it is sufficient that $y \in Y$. The operator $\suff{R}$ has the following decisive properties for all $Y,Z \subseteq X$:
\begin{xalignat}{2}
\suff{R}(\emptyset) &= X, &&\text{Co-normality,} \\
\suff{R}(Y \cup Z) &= \suff{R}(Y) \cap \suff{R}(Z), && \text{Co-additivity.}
\end{xalignat} 
Generalizing $\suff{R}$ to Boolean algebras we arrive at the following definition: For a Boolean algebra $B$ a mapping $g: B \to B$ is called a \emph{sufficiency operator} if it is satisfies for all $a,b \in B$
\begin{align}
g(\zero) & = \one, \\
g(a+b) &= g(a) \cdot g(b).
\end{align}
The structure $\klam{B,g}$ is called a \emph{sufficiency algebra}. A sufficiency operator is in some sense  a complementary counterpart to the possibility operator. It also is the algebraic counterpart of the window modality, which can be traced back to the works of \citet{Humberstone-IW}, \citet{gpt87}, \citet{gor88} and also \citet{vanBenthem-MDL} in the framework of deontic logic. The theory of sufficiency algebras in conjunction with possibility algebras was refined and developed further by \citet{do_mixalg}, \citet{dot_mixed}, and \citet{dgm23,dgm26}. 

Suppose that $g\colon B \to B$ is a mapping and define $g^\star$ by $g^\star(a) \df -{g(a)}$. 

\begin{theorem} \label{thm:necsuff} \cite{do_baro}
$g$ is a sufficiency operator \tiff $g^\star:B \to B$ is a possibility operator.
\end{theorem}
The \emph{canonical structure} $\Cs(\B)$ of a sufficiency algebra $\B = \klam{B,g}$ is the structure $\klam{X,R_g}$ where $X$ is the set of ultrafilters of $B$ and $R_g$ is the relation on $X$ defined by
\begin{gather}\label{def:CfSuff}
F \mathrel{R_g} G \Iffdf g[G] \cap F \neq \z.
\end{gather}
Conversely, a \emph{sufficiency frame}   is a structure $\klam{X,R}$ where $X$ is a nonempty set and $R$ is a binary relation on $X$. The notation emphasizes that the complex algebra of a sufficiency frame defined below is a sufficiency algebra. Its complex algebra $\Cm(X)$ is the structure $\klam{2^X, \suff{R}}$, where $2^X$ is the powerset algebra of $X$, and $\suff{R}\colon 2^X \to 2^X$ is the mapping defined by $\suff{R}(Y) \df \set{x \in X: R(x) \subseteq Y}$. 

We now have the duality results for sufficiency structures.

\begin{theorem}\label{thm: repsuff1}
\begin{enumerate}
\item\cite[Proposition 3.3.2]{orr15}  The complex algebra of a sufficiency frame is a sufficiency algebra.
\item\cite[Proposition 3.3.3]{orr15} The canonical structure of a sufficiency algebra is a sufficiency frame.
\end{enumerate}
\end{theorem}

\begin{theorem}\label{thm: repsuff2}\cite[Proposition 3.3.6)]{orr15} 
\begin{enumerate}
\item Every sufficiency algebra can be embedded into the complex algebra of its canonical structure by the mapping $h$ of Theorem \ref{thm:monad}.
\item Every sufficiency frame can be embedded into the canonical structure of its complex algebra. 
\end{enumerate}
\end{theorem}
While $\poss{R}$ and its dual $\nec{R}$ talk about properties of the relation $R$, the sufficiency operator can express properties of the complement of $R$, since
\begin{gather}\label{necsuff}
\suff{R}(Y) = \nec{X^2 \setminus R}(-Y).
\end{gather}
For example, $\nec{R}$ can express reflexivity, while $\suff{R}$ can express irreflexivity. In the current context we are considering approximations spaces, so the relation $R$ in question will be an equivalence relation $\theta$, and $\suff{\theta}(Y)$ is in a sense the lower approximation of $Y$ with respect to the complement of $\theta$. 

Another construction is worthy of mention in connection with the complement of an equivalence relation. A binary relation $R$ on $X$ is called \emph{co-transitive}, if for all $x,y \in X$, $x \mathrel{R} y $ implies that $x \mathrel{R} z$ or $z \mathrel{R} y$. It is not hard to see that $R$ is co-transitive \tiff its relational complement is transitive. A \emph{diversity relation} on $X$ is a binary relation $R$ which satisfies
\begin{align}
&\tag{FDiv1}R \text{ is irreflexive.} \\
&\tag{FDiv2}R \text{ is symmetric.} \\
&\tag{FDiv3}R \text{ is co-transitive.} 
\end{align}
The diversity relations on $X$ are exactly the complements of equivalence relations, and thus they are of interest for investigating structures arising from rough sets. A \emph{diversity frame} is a structure $\klam{X,R}$, where $X$ is a nonempty set and $R$ is a diversity relation. The algebraic counterpart are \emph{diversity algebras} which are sufficiency algebras $\klam{B,g}$ which satisfy
\begin{align}
\tag{Div1} g(a) &\leq -a. \\
\tag{Div2} a &\leq g(g(a)). \\
\tag{Div3} g(a) &\leq g(-g(a)).
\end{align} 
Complex algebras or diversity frames and canonical structures are defined as for sufficiency algebras and sufficiency frames. Following Theorems \ref{thm: repsuff1} and \ref{thm: repsuff2} we obtain the duality results:
\begin{theorem}\label{thm: repdiv1}
\begin{enumerate}
\item\cite[Proposition 4.6.1]{orr15}  The complex algebra of a diversity frame is a diversity algebra.
\item\cite[Proposition 4.6.2]{orr15} The canonical structure of a diversity algebra is a diversity frame.
\end{enumerate}
\end{theorem}
\begin{theorem}\label{thm: repdiv2}\cite[Theorem 4.6.3]{orr15} 
\begin{enumerate}
\item Every diversity algebra can be embedded into the complex algebra of its canonical structure by the mapping $h$ of Theorem \ref{thm:monad}.
\item Every diversity frame can be embedded into the canonical structure of its complex algebra. 
\end{enumerate}
\end{theorem}
Since the complement of $R$ is an equivalence relation $\theta$, we see from \eqref{necsuff} that $\suff{R}(Y)$ is the lower approximation of $-Y$ with respect to $\theta$.

\section{Algebras of rough sets}

A second method to obtain algebraic structures from Pawlak's approximation spaces is to provide the collection of rough sets itself with an algebraic structure. Let us take again an approximation space $\klam{X,\theta}$ as a starting point. In Section \ref{sec:modal} we have looked at the Boolean power set algebra $2^X$ and augmented it by a possibility operator. In this section we shall consider the collection $2^X_\theta$ of rough sets of $\klam{X,\theta}$ and define appropriate operators on $2^X_\theta$. We start with the lattice operations:
\begin{align}
\klam{\underline{Z},\overline{Z}} \lor \klam{\underline{Y},\overline{Y}} &\df \klam{\underline{Z} \cup \underline{Y},\overline{Z} \cup \overline{Y}}, \label{rJoin} \\
\klam{\underline{Z},\overline{Z}} \land \klam{\underline{Y},\overline{Y}} &\df \klam{\underline{Z} \cap \underline{Y},\overline{Z} \cup \overline{Y}}, \label{rMeet} \\
\zero &\df \klam{\z,\z}, \label{r0} \\
\one &\df \klam{X,X}. \label{r1}
\end{align}
It is well known, that with these operations $2^X_\theta$ becomes a bounded distributive lattice \cite{iwi87}. Various forms of negations were defined on $2^X_\theta$ which lead to different classes of algebras among which very strong connections hold. For an in depth investigation of these we invite the reader to consult \cite{pag_roughlog}. In the present article we will consider two of these, the \emph{regular double Stone algebras} and the \emph{De Morgan} algebras. 

All these algebras of rough sets are based on bounded distributive lattices, so we first describe the discrete duality for bounded distributive lattices based on Stone's construction \cite{stone_DL}. Suppose that $L$ is a bounded distributive lattice. The \emph{canonical structure} $\Cf(L)$ of $L$ is the partially ordered set $\klam{\Prim(L), \subseteq}$. Conversely, if $\klam{X, R}$ is a partial order, its \emph{complex algebra} $\Cm(X)$ is the structure $\klam{L_X, \cup, \cap, \z, X}$, where $L_X \df \set{Y \subseteq X: \ua{Y} = Y}$. Note that the universe of $\Cf(\Cm(X))$ is the set of prime filters of $\Cm(X)$, and that the universe of $\Cm(\Cf(L))$ is the power set lattice of $\Prim(L)$. 

We now have the discrete duality for bounded distributive lattices and ordered frames:
\begin{theorem}\label{thm:DDLat}
\begin{enumerate}
\item\cite{stone_DL} For each ordered frame $\klam{X,\leq}$ $\Cm(X)$ is a bounded distributive lattice, and the mapping $h: L \into 2^{\Prim(L)}$ defined by $h(a) \df \set{F \in \Prim(L): a \in F}$ is an embedding of $L$ into $\Cm(\Cf(L))$. 
\item\cite{orr15} For each bounded distributive lattice $L$ $\Cf(L)$ is a partially ordered set, and the mapping  $k: X \to \Cf(L_X)$ defined by $k(x) \df \set{A \in \Cm(X): x \in A}$ is a frame embedding of $X$ into $\Cf(\Cm(X))$.  \qed
\end{enumerate}
\end{theorem}
In the sequel we will assume these embeddings for extensions of bounded distributive lattices and ordered frames.

\section{Regular double Stone algebras}\label{sec:rdsa}

 Following \citet{iwi87}, \citet{pp88}, and \citet{com93} we define the \emph{full algebra of rough sets} over the approximation space $\klam{X,\theta}$ as the structure $\mathcal{P}_\theta(X) \df \klam{2^X_\theta, \lor, \land, {}^*, {}^+, \zero, \one}$, where $\zero \df \klam{\z,\z}, \one \df \klam{X,X}$, and
\begin{align}
\klam{\underline{Y},\overline{Y}}^* &\df \klam{Y \setminus \overline{Y}, Y \setminus \overline{Y}}, \label{c*}\\
\klam{\underline{Y},\overline{Y}}^+ &\df \klam{Y \setminus \underline{Y}, Y \setminus \underline{Y}}, \label{c+}
\end{align}
where the approximations are taken with respect to $\theta$. An \emph{algebra of rough sets} is a subalgebra of some full algebra of rough sets. It turns out that the class of algebras of rough sets has properties well known from lattice theory since the late 1960s \cite{var68}, which we shall now describe. A \emph{double Stone algebra} (DSA) $\langle L, +, \cdot, {}^*, {}^+, \zero, \one\rangle$ is an algebra of type $\langle 2,2,1,1,0,0\rangle $ such that
\begin{enumerate}
\item $\langle L, + , \cdot, \zero, \one\rangle $ is a bounded distributive lattice,
\item $a^*$ is the pseudocomplement of $a$, i.e.
$
b \leq a^* \Leftrightarrow b \cdot a = \zero,
$
\item $a^+$ is the dual pseudocomplement of $a$, i.e.
$
b \geq a^+ \Leftrightarrow b + a = \one,
$
\item $a^* + a^{**} = \one$, $a^+ \cdot a^{++} = \zero$.
\end{enumerate}

A double Stone algebra $L$ is called \emph{regular} (RDSA), if it additionally satisfies 
\begin{gather}
\tag{M}\label{M} a^* = b^* \tand a^+ = b^+ \timplies a = b.
\end{gather}
It is well known that the class of RDSAs is equational \cite{var69}. Condition \eqref{M} shows that each $a \in L$ is uniquely determined by the pair $\klam{a^{++}, a^{**}}$ just as every rough set is determined by some $\klam{\underline{X}, \overline{X}}$. RDSAs were studied, among others, by \citet{var68,var69,var72} and \citet{kat73,kat74}. 

\citet{moi40,moi72} proposed a class of algebras referred to as \emph{3--valued {\L}ukasiewicz algebras}, which are closely related to {\L}ukasiewicz's 3-valued logic.  It can be shown that regular double Stone algebras and {\L}ukasiewicz algebras can be interdefined,  and condition \eqref{M} is named in his honour by \citet{var69}. There is also a close connection to Nelson algebras \cite{pag_roughlog}. A logic for rough sets based on regular double Stone algebras and their representation developed by \citet{kat74} is presented in \cite{id_roughlog}.

There is a close relationship of algebras of rough sets to RDSAs, in particular, each RDSA is representable as an algebra of rough sets:: 
\begin{theorem}\label{rhm:rsdaalg}
\begin{enumerate}
\item Each algebra of rough sets is an RDSA \cite{pp88,com93}.
\item Every regular double Stone algebra is isomorphic to an algebra of rough subsets of an approximation space \cite{com93}. \qed
\end{enumerate}
\end{theorem}
The main task to establish a discrete duality for RDSAs is to find a frame condition on an ordered frame $\klam{X,\leq}$ which guarantees that $\Cm(X)$ is an RDSA.  The decisive condition relates regularity to the ordered set of prime ideals of $L$:
\begin{lemma}\label{lem:reg} \cite{var68}
Let $L$ be a double Stone algebra. Then, the following statements are equivalent:
\begin{align}
& L \text{ is regular}. \label{reg1} \\
&\text{$\klam{\Prim(L), \subseteq}$ consists of chains of length at most two. \hfill{\qed}} \label{reg_chain}
\end{align}
\end{lemma}
We will use \eqref{reg_chain} to obtain a suitable frame condition. A partially ordered set. $\klam{X,\leq}$ is called an \emph{RDSA frame}, if it satisfies
\begin{gather}
\tag{RDS}\label{rds1} (\forall x,y)[(x \leq y \Implies x = y) \tor ((\forall z)[x \leq z \Implies z = y])].
\end{gather}
\eqref{rds1} says that above each $x \in X$ there is at most one element different from $x$, which implies that every order component of $\klam{X,\leq}$ is a chain of length at most two. We let $\max(X)$ be the set of maximal points of $X$, and $\min(X)$ the set of its minimal points.  The \emph{complex algebra of an RDSA frame $\klam{X,\leq}$} is the structure $\Cm(X) = \klam{L_X, \cap, \cup, \z, {}^*, {}^+, X}$, where $L_X$ is as in Theorem \ref{thm:DDLat}, and or $Y \subseteq X$
 \begin{align}
 Y^* &= - \da(\max(X) \cap Y), \tand \\
 Y^+ &= - \ua(\min(X) \cap Y).
 \end{align} 
It turns out that together with Theorem \ref{thm:DDLat} this is enough for the discrete duality:
\begin{theorem}\label{thm:rdsa1}\cite[Theorem 5.2]{do_stone} 
\begin{enumerate}
\item If $L$ is a regular double Stone algebra, then $\Cf(L)$ is an RDSA frame.
\item If the partially ordered set $\klam{X,\leq}$ is an RDSA frame, then $\Cm(X)$ is a regular double Stone algebra.
\end{enumerate}
\end{theorem}

\begin{theorem}\label{thm:rdsa2}\cite[Theorem 3.3]{do_stone} 
\begin{enumerate}
\item Every RDSA is embeddable into the complex algebra of its canonical structure.
\item Every RDSA frame is embeddable in to the canonical structure of its complex algebra.
\end{enumerate}
\end{theorem}

\section{De Morgan algebras}

A different negation $\neg$ can by defined on $2^X_\theta$ by setting
\begin{gather}
\neg\klam{\underline{Y}, \overline{Y}} \df \klam{\underline{-Y}, \overline{-Y}}.
\end{gather}
It turns out that the resulting structure is an instance of the well known De Morgan algebras \cite{iwi87,bc04}. A \emph{De Morgan algebra}\footnote{De Morgan algebras were called \emph{quasi-Boolean algebras} in the Polish school of logic \cite{br57}.} is a bounded distributive lattice $L$ augmented by a unary operator $\neg$ on $L$ which satisfies
\begin{align}
\tag{DeM1}\label{DeM1}\neg\neg a & = a, \\
\tag{DeM2}\label{DeM2} \neg(a+b) &= \neg a \cdot \neg b.
\end{align}

A \emph{De Morgan frame} is a relational structure $\klam{X,\leq,N}$ where $\klam{X,\leq}$ is an ordered frame, and $N\colon X \to X$ is a function which satisfies
\begin{align}
\tag{FDeM1}\label{FDeM1} N(N(x)) &=  x, \\
\tag{FDeM2}\label{FDeM2} x \leq y &\Implies N(y) \leq N(x).
\end{align}
The complex algebra of a De Morgan frame $X$ is the structure $\Cm(X) \df \klam{L_X, \neg_N, \z, X}$, where $L_X$ is the complex algebra of the ordered frame $\klam{X,\leq}$ as in Theorem \ref{thm:DDLat}, and
\begin{gather}
\neg_N(Y) \df - N[Y]
\end{gather}
for all $Y \subseteq X$. Conversely, the canonical structure of a De Morgan algebra $\klam{L, \neg}$ is the structure $\Cs(L) \df \klam{X_L, N_\neg}$, where $X_L$ is the canonical frame of the lattice reduct of $L$, and 
\begin{gather}
N_\neg(F) \df L \setminus \neg[F].
\end{gather}
for all $F \in \Prim(L)$. We now have the discrete duality:
\begin{theorem}\label{thm: DeM}\cite[Section  9.8]{orr15} 
\begin{enumerate}
\item The complex algebra of a De Morgan frame is a De Morgan Algebra.
\item The canonical structure of a De Morgan algebra is a De Morgan frame.
\item Every De Morgan algebra is embeddable into the complex algebra of its canonical structure.
\item Every De Morgan frame is embeddable in to the canonical structure of its complex algebra.
\end{enumerate}
\end{theorem}

\section{Rough relation algebras}

Pawlak's original approach to model incomplete information was to take sets as the basic entity. However, sets can themselves have an underlying structure or a special form.   Given an approximation space $\klam{X,\theta}$ we can lift $\theta$ to a relation $\theta^2$  on $X^2$ by defining 
\begin{gather}\label{def:R2}
\klam{x,y}\mathrel{\theta^2} \klam{x',y'} \Iffdf x \mathrel{\theta} x' \tand y \mathrel{\theta} y'.
\end{gather}
Clearly, $\theta^2$ is an equivalence relation, and we can consider the approximation space $\klam{X^2,\theta^2}$ \cite{paw81d}. A \emph{rough relation} with respect to $\theta^2$ is a pair $\klam{\underline{R}, \overline{R}}$ where $R \subseteq X^2$ and the approximations are taken with respect to $\theta^2$; we denote the collection of all rough relations on $X$ with respect to $\theta$ by $\rel_\theta(X)$.  Its associated \emph{full algebra of rough relations} is the structure $\mathcal{P}_{\theta^2}(X) \df \klam{\rel_\theta(X), \lor, \land, {}^*, {}^+, \zero, \one}$, where $\zero \df \z, \one \df X \times X$, the lattice operations are defined in \eqref{rJoin} and \eqref{rMeet}, and the pseudocomplements are defined in \eqref{c*} and \eqref{c+}.

Since our objects are binary relations we have the classical relational operations at our disposal to enhance the expressive power of the language, namely, relational composition, relational converse , and the identity relation as a new constant. These are defined on the set $\rel(X)$ of binary relations on $X$ as follows:
\begin{xalignat*}{2}
R \comp S &\df \set{\klam{x,y}: (\exists z)[x\mathrel{R}z \text{ and } z\mathrel{S}y]}, && \text{Composition}, \\
\conv{R} &\df \set{\klam{y,x}: x\mathrel{\theta} y}, &&\text{Converse}, \\
\id & \df \set{\klam{x,x}: x \in X}, && \text{Identity}.
\end{xalignat*}

To approximate binary relations on $X$ these relational operators can be generalized to act on rough relations \cite{com93}. With some abuse of language we use the same notation for operators on rough relations as for the operators on binary relations:
\begin{xalignat*}{2}
\klam{\underline{R}, \overline{R}} \comp \klam{\underline{S}, \overline{S}} &\df \klam{\underline{R} \comp \underline{S}, \overline{R} \comp \overline{S}} && \text{Rough composition} \\
\conv{\klam{\underline{R}, \overline{R}}} &\df \klam{\underline{\conv{R}}, \overline{\conv{R}}} && \text{Rough converse} \\
\id &\df \klam{\theta,\theta} && \text{Rough identity}
\end{xalignat*}
The structure $\klam{\rel_\theta(X), \lor, \land, {}^*, {}^+, \z,X \times X, \comp, \conv{}, \id}$ is called the \emph{full algebra of rough relations over $\klam{X^2,\theta^2}$)}. An \emph{algebra of rough relations} is a subalgebra of some full algebra of rough relations.

Rough relation algebras were introduced by \citet{com93} and further investigated in \citet{id_roughrel,dw_r2a}. These are intended to serve as an abstract counterpart to algebras of rough relations. A \emph{rough relation algebra} (R2A) is a structure $\klam{L,+, \cdot, {}^*, {}^+,\zero,\one , \comp\;, \conv{},\id}$ of type $\klam{2,2,1,1,0,0,2,1,0}$ satisfying the following axioms:
\begin{align}
\tag{R2A$_0$}\label{a0}  & \text{$\klam{L, +, \cdot,{}^*,{}^+,\zero, \one}$ is a regular double Stone algebra.} \\
\tag{R2A$_1$}\label{a1} & \text{$\klam{L, \comp\;, \id }$ is a semigroup with identity $\id $.} \\
\tag{R2A$_2$}\label{a2} & a \comp (b+c) = (a \comp b) + (a \comp c),  \quad (b+c) \comp a = (b \comp a) + (c \comp a). \\
\tag{R2A$_3$}\label{a3} & \conv{\conv{a}} = a. \\
\tag{R2A$_4$}\label{a4} & \conv{(a+b)} = \conv{a} + \conv{b}. \\
\tag{R2A$_5$}\label{a5} & \conv{(a\comp b)} = \conv{b} \comp \conv{a}. \\
\tag{R2A$_6$}\label{a6} & \conv{a} \comp (a \comp b)^* \leq b^*. \\
\tag{R2A$_7$}\label{a7} & (a^* \comp b^*)^{**} = a^* \comp b^*. \\
\tag{R2A$_8$}\label{a8} & \id ^{**} = \id .
\end{align}
These statements generalize the axioms of Tarski's relation algebras \cite{tar41} by replacing the Boolean part by a regular double Stone algebra and an adjusted set of axioms. Unlike regular double Stone algebras, not every R2A is representable as a subdirect product of algebras of rough relations. A condition for representability is given in \cite[Proposition 4.6]{id_roughrel}.

To find an appropriate class of frames for the class of R2As we will use the discrete duality for regular double Stone algebras of Theorem \ref{thm:rdsa2}. As we have three additional operators, namely, $\comp$\;, $\conv{}$, and $\id$ of arity, respectively, $2$, $1$, and $0$, we require corresponding relations of rank $3$, $2$ and $1$. This leads to the following definition: 

A \emph{rough relation frame} is a structure $\X = \klam{X, \leq, R, f, I}$ where $\leq$ is a partial order on $X$ which is the disjoint union of chains of length at most two, $R$ is a ternary relation on $X$, $f:X \to X$ a function, and $\z \neq I \subseteq X$. If $x \in X$, we denote the minimal element below $x$ by $\underline{x}$, and the maximal element above $x$ by $\overline{x}$; observe that $x \in \set{\underline{x}, \overline{x}}$. We postulate the following axioms:
\begin{align}
\tag{R2FA$_0$}\label{ra0}  & \text{$R(x,y,z)$, $x' \leq x$, $y' \leq y$, $z \leq z'$ $\Implies$ $R(x',y',z')$.} \\
\tag{R2FA$_1$}\label{ra1} & \text{$x \leq y$ $\Implies$ $f(x) \leq f(y)$} \\
\tag{R2FA$_2$}\label{ra2} & \text{$I$ is $\leq$--closed and $\geq$--closed.} \\
\tag{R2FA$_3$}\label{ra3} & \text{ $R(x,y,z) \tand R(z,v,w) \Implies (\exists u)[R(x,u,w) \tand R(y,v,u)]$.} \\
\tag{R2FA$_4$}\label{ra4} & \text{ $R(x,y,z) \tand R(v,z,w) \Implies (\exists u)[R(u,y,w) \tand R(v,x,u)]$.} \\
\tag{R2FA$_5$}\label{ra5} & \text{$f(f(x)) = x$.} \\
\tag{R2FA$_6$}\label{ra6} & \text{$f(\overline{x}) = \overline{f(x)}, f(\underline{x}) = \underline{f(x)}$.} \\
\tag{R2FA$_7$}\label{ra7} & R(x,y,z) \Implies R(f(x),z,\overline{y}). \\
\tag{R2FA$_8$}\label{ra8} & R(x,y,z) \Implies R(z,f(y),\overline{x}). \\
\tag{R2FA$_9$}\label{ra9} & R(x,y,z) \Implies R(\underline{x}, \underline{y},\underline{z}). \\
\tag{R2FA$_{10}$}\label{ra10} & x \leq y \Iff (\exists z)[z \in I \tand R(x,z,y)]. \\
\tag{R2FA$_{11}$}\label{ra11} & z \leq y \Iff (\exists x)[x \in I \tand R(x,z,y)].
\end{align}

The complex algebra of $X$ has the set $L_X$ of increasing subsets of $\klam{X,\leq}$ as universe. The operations on $L_X$ are defined as follows: If $Y \in L_X$, then
\begin{align*}
Y^* &\df \set{y: \ua \set{y} \cap Y = \z}, \\
Y^+ &\df \set{y: \da \set{y} \cap -Y \neq \z}.
\end{align*}
Then, $Y^*$ is the largest increasing subset contained in $X \setminus Y$, and $Y^+$ is the smallest increasing subset containing $X \setminus Y$. 

We now can state the discrete duality for rough relation algebras:

\begin{theorem} \cite{do13} Let $L$ be a rough relation algebra, and  $\X = \klam{X, \leq, R, f, I}$  be a rough relation frame.
\begin{enumerate}
\item The canonical frame of a rough relation algebra is a rough relation frame.
\item The complex algebra of a rough relation frame is a rough relation algebra.
\item The Stone embedding $h\colon L \to \Cm\Cs(L)$ defined by $h(a) \df \set{F \in Prim(L): a \in F}$ is an embedding of rough relation algebras.
\item The mapping $k\colon X \to \Cs\Cm(X)$ defined by $k(x) = \set{A \in L_X: x \in A}$ is a frame embedding. 
\end{enumerate}
\end{theorem}

\section{Summary and outlook}

In this paper we have given an overview of discrete dualities for classes of algebras and classes of frames chosen according to a twofold criterion. The chosen classes of algebras provide a characterisation of Pawlak’s approximation spaces, and the chosen classes of frames expose a distinctive nature of relational systems which are the counterparts of those algebras. Together they provide a view for strategies of defining formalised theories for characterisation of various approaches to approximate missing information. In future work we aim to exhibit discrete dualities for deterministic and indeterministic information systems in the sense of \citet{lipski79}. 

\section{References}
\renewcommand*{\refname}{}
\vspace{-10mm}


\end{document}